\documentclass[sigconf]{acmart}

\usepackage{kotex}
\usepackage{comment}

\AtBeginDocument{%
  \providecommand\BibTeX{{%
    \normalfont B\kern-0.5em{\scshape i\kern-0.25em b}\kern-0.8em\TeX}}}

\setcopyright{acmlicensed}
\setcopyright{acmcopyright}
\copyrightyear{2022}
\acmYear{2022}
\acmConference[CHI '22 Extended Abstracts]{CHI Conference on Human Factors in Computing Systems Extended Abstracts}{April 29-May 5, 2022}{New Orleans, LA, USA}
\acmBooktitle{CHI Conference on Human Factors in Computing Systems Extended Abstracts (CHI '22 Extended Abstracts), April 29-May 5, 2022, New Orleans, LA, USA}
\acmPrice{15.00}
\acmDOI{10.1145/3491101.3519854}
\acmISBN{978-1-4503-9156-6/22/04}

\begin{document}

\title[AI-assisted Emotional Support in OHMCs]{Exploring the Effects of AI-assisted Emotional Support Processes in Online Mental Health Community}

\author{Donghoon Shin}
\affiliation{%
  \institution{Seoul National University}
  \city{Seoul}
  \country{Korea}}
\email{ssshyhy@snu.ac.kr}

\author{Subeen Park}
\affiliation{%
  \institution{Seoul National University}
  \city{Seoul}
  \country{Korea}}
\email{psbpsu@snu.ac.kr}

\author{Esther Hehsun Kim}
\affiliation{%
  \institution{Seoul National University}
  \city{Seoul}
  \country{Korea}}
\email{ehk@snu.ac.kr}

\author{Soomin Kim}
\affiliation{%
  \institution{Seoul National University}
  \city{Seoul}
  \country{Korea}}
\email{soominkim@snu.ac.kr}

\author{Jinwook Seo}
\affiliation{%
  \institution{Seoul National University}
  \city{Seoul}
  \country{Korea}}
\email{jseo@snu.ac.kr}

\author{Hwajung Hong}
\affiliation{%
  \institution{KAIST}
  \city{Daejeon}
  \country{Korea}}
\email{hwajung@kaist.ac.kr}

\renewcommand{\shortauthors}{Shin, et al.}

\begin{abstract}
    Social support in online mental health communities (OMHCs) is an effective and accessible way of managing mental wellbeing. In this process, sharing emotional supports is considered crucial to the thriving social supports in OMHCs, yet often difficult for both seekers and providers. To support empathetic interactions, we design an AI-infused workflow that allows users to write emotional supporting messages to other users' posts based on the elicitation of the seeker's emotion and contextual keywords from writing. Based on a preliminary user study (N = 10), we identified that the system helped seekers to clarify emotion and describe text concretely while writing a post. Providers could also learn how to react empathetically to the post. Based on these results, we suggest design implications for our proposed system.

\end{abstract}



\begin{teaserfigure}
  \centering
  \includegraphics[width=.8\textwidth]{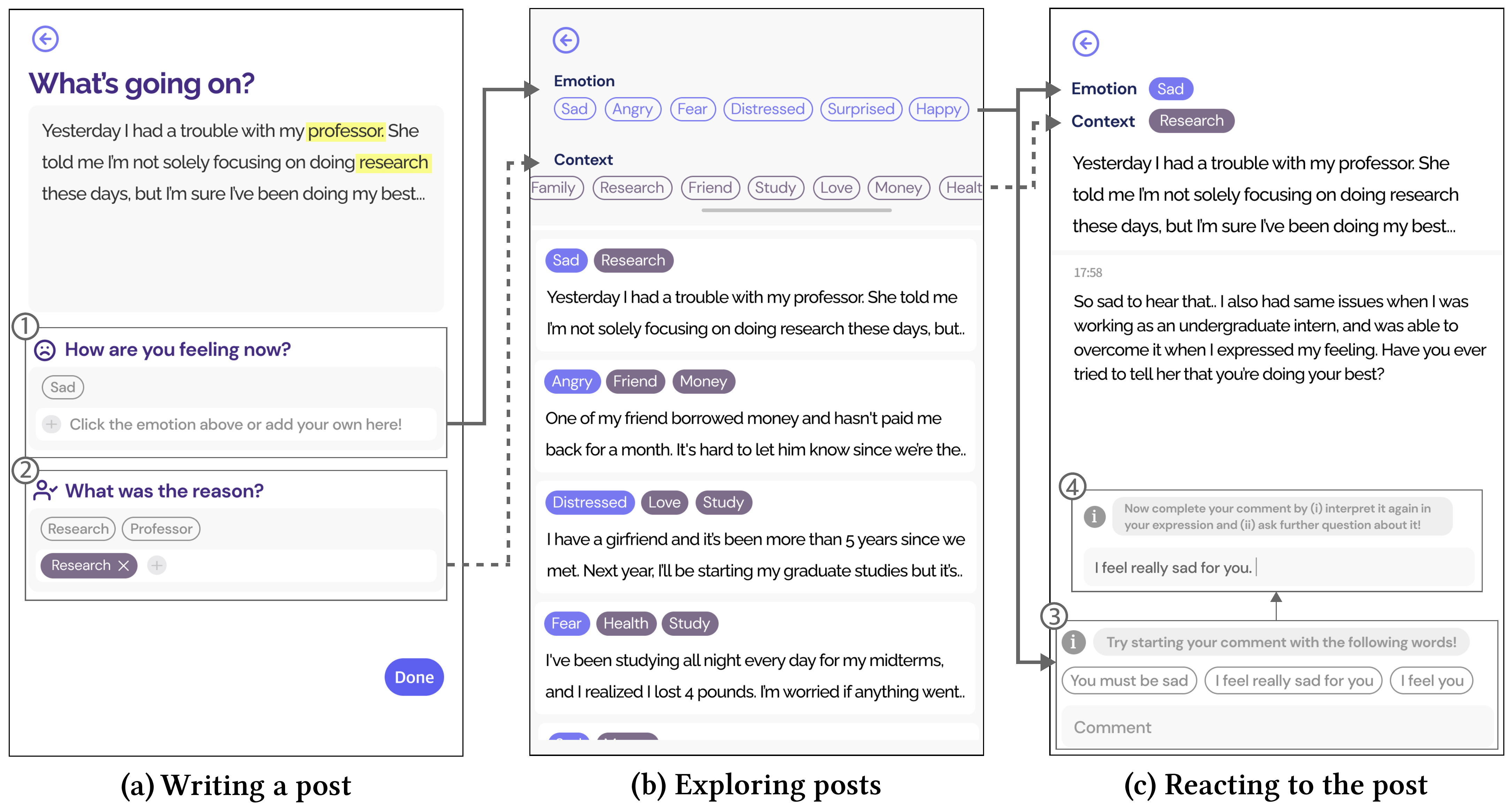}
  \caption{Keyscreens of our system we designed and examined. Once the seeker writes down the post, the system automatically detects (a)-\textcircled{1} emotion and (a)-\textcircled{2} contextual keywords, where the user can select among them or newly add one. (b) By filtering posts using the metadata from (a), providers can explore posts among the list of posts. Once clicking the post, providers can start reading it and (c)-\textcircled{3} commenting empathetically based on three phrase recommendations for emotional reaction, followed by (c)-\textcircled{4} the prompts asking users to leave interpretation and exploration of the post. The interface elements were translated from Korean}
  \label{fig:teaser}
\end{teaserfigure}

\maketitle

\section{Introduction}

Online mental health communities (OMHCs) have become a prevalent medium of promoting mental wellness through collaborative interactions among people~\cite{sharma2018mental}. For instance, users in multiple mental health sub-Reddits benefit from actively sharing their challenging experiences and gaining feedback from their peers online~\cite{sharma2018mental}. By posting their experiences and concerns and reacting to them, individuals in OMHCs can actively share emotional and informational support with people with similar conditions~\cite{cutrona1994social}.

Here, ensuring active emotional support-sharing processes among users is particularly crucial to maintaining thriving OMHCs~\cite{elliott2011empathy}, yet often considered challenging~\cite{prescott2019young}. For providers, it is often burdensome and overwhelming to be emotionally engaged~\cite{highet2004much, barney2009exploring}, leading to their dropouts~\cite{de2015feels}. Plus, since most of the existing OMHCs are text-based~\cite{hoermann2017application}, providers have difficulty in converting their empathetic thoughts in a text-based form~\cite{gibson2016deep, perez2017understanding}. Similarly, it is equally important, yet difficult, for seekers to ensure that they disclose their experiences concretely to help providers better understand seeker's experiences and react emotionally~\cite{altman1973social, yang2019channel}.

In line with such concerns, previous studies have devised theoretical guidance to support individuals to learn how to react emotionally in text-based settings~\cite{sharma2020computational}. On top of these, recent studies have begun to feature the role of AI in supporting emotional reaction processes. For example, Peng et al. designed MepBots, a writing assistance that aids empathetic reaction processes~\cite{peng2020exploring}. Still, such an approach is limited in that the quality of the seeker's posts (e.g., concreteness) is also a crucial factor that affects the emotional supports~\cite{li2016sunforum, yang2019channel}. Thus, rather than merely supporting the empathy-providing process, understanding the holistic workflow - from writing the post (seeker's side) to navigating posts and reacting to such seekers (provider's side) - is important.

Thus, we aim to explore the feasibility of AI in augmenting the overall workflow of empathetic communication by considering both seeker and provider sides in OMHCs. Specifically, we designed and evaluated an AI-infused mental health community app, which supports the scaffolded interactions (writing posts, exploring posts, and reacting to posts) specialized to facilitate empathetic communication in OMHCs by detecting the emotion and contextual keywords and recommending appropriate triggers/prompts to assist provider's reaction. From the preliminary user study with 10 participants, we identified the feasibility and areas of improvement for enhancing emotional support processes in AI-infused OMHCs.






\section{Related Work}

\subsection{Online Mental Health Community}

Online mental health communities (OMHCs) are the prevalent medium of exchanging social support with peers experiencing similar challenges, providing high accessibility and anonymity~\cite{o2017design, sharma2018mental, peng2021effects, li2016sunforum, naslund2016future}. To date, several studies have explored the dynamics of peer supports in OMHCs or mental health-related discussions, such as relationship between user satisfaction and their community knowledge~\cite{peng2021effects}, cross-cultural variances of OMHCs~\cite{pendse2019cross}, and motivations and practices of mental health discussions in Twitch~\cite{uttarapong2022twitch}.

For such OMHCs to thrive and individuals to improve mental health (e.g., reducing depression/anxiety), it is known that active sharing (e.g., sharing experiences, expressing emotions, and restructuring maladaptive cognition) are particularly crucial~\cite{lepore2002expressive, kushner2020bursts, morris2015efficacy, pfeiffer2011efficacy, pruksachatkun2019moments, radcliffe2007written}. Specifically, previous works suggested that people were more satisfied when their received support matches the needed support~\cite{wolff2013physical, vlahovic2014support}, in the form of informational or emotional support. These are known to increase both members’ commitment and satisfaction when they sought and received the matched support~\cite{yang2017commitment, peng2021effects}. Here, emotional supports are particularly known to lead support-seeker to express more satisfaction when receiving emotional support from the reaction~\cite{peng2021effects}. Thus, to guide long-lasting and thriving OMHC environments, it is crucial to ensure emotional supports between seekers and providers.

However, sharing emotional supports in OMHCs is especially challenging for both support seekers and providers~\cite{gibson2016deep, perez2017understanding}. For instance, providers often find it difficult to convert their empathetic thoughts in a text-based form~\cite{gibson2016deep, perez2017understanding}. Similarly, it is difficult for seekers to disclose their experiences concretely, which helps providers better understand their experiences and react emotionally~\cite{altman1973social, yang2019channel}. As such, researchers have begun to devise systems that support the writing process of seekers and providers~\cite{wu2019design, peng2020exploring}. However, they were limited in that the system (i) assessed posts based on the quality, which may discourage and negatively affect users~\cite{wu2019design}, or (ii) only focused on the provider's side~\cite{peng2020exploring}.

\subsection{AI-mediated Communication}
Advances in artificial intelligence (AI) and machine learning (ML) have transformed the paradigm of how people communicate and collaborate, and changed the pattern of forming interpersonal relationships. While traditional computer-mediated communication (CMC) focuses on the interactions in which communication is transmitted and mediated via technologies, AI-Mediated Communication (AI-MC) assumes that communication is modified, amplified, and even generated by a computational agent~\cite{hancock2020ai}. For example, recommendation algorithms used in social networking online dating applications determine who we associate and converse with.

Various methods have been proposed to enhance interpersonal relationships using AI and ML technology. The most representative case would be recommending appropriate responses. Human-AI hybrid conversational system was developed for reply suggestions~\cite{gao2021evaluating}. Robertson et al.~\cite{robertson2021can} proposed a system that classifies problematic email reply suggestions by analyzing email content (e.g., semantic, tonal coherence) and social context (e.g., communication dynamics). Furthermore, an AI-infused chatbot agent was proposed to recommend common interests between the strangers based on their social media data~\cite{shin2021blahblahbot}.

In line with previous studies of using the assists of AI to enhance interpersonal communication, this work premises that technology could support the mental health community by augmenting interpersonal relationships between community members~\cite{gui2017investigating}. Specifically, we design and develop a system that aids emotional support workflow in OMHCs by harnessing AI to (i) detect/recommend emotion/contextual keywords for seekers and (ii) recommend triggers/scaffolds of empathetic emotional reactions for providers. Based on the preliminary user study with 10 users, we discuss enhancements on AI-infused emotional support workflow across seekers and providers.

\section{Prototype Design} \label{mvp}

\subsection{Overall Concept and Workflow}

The goal of our system is to support empathetic peer-support workflow in OMHCs. Here, in addition to identifying the \textit{emotion} of the seeker's post, it is also important to ensure that providers understand the overall \textit{contextual information} (e.g., place, reason, subject in conflict) to provide more detailed and empathetic response~\cite{pfeil2007patterns}. Thus, we decided to let users elicit and leverage (i) emotion and (ii) contextual information through their exchange, with the aid of AI.

\subsection{System Design}

In OMHCs, supports between providers and seekers are often shared as a form of \textit{post}, a text-based unit that basically contains content written by the seeker and its comments from the providers. Specifically, this process can be conceptualized with the following interactions: (i) seekers post their emotional contexts, (ii) providers explore such posts in the community, and they (iii) react to the post~\cite{gui2017investigating, peng2020exploring, sharma2018mental, peng2021effects}. Thus, we designed our proposed system by following three interactions:

\subsubsection{Writing a post}

First, the user is asked to freely describe their personal experiences that required emotional support (Figure \ref{fig:teaser}-(a)). Once the user enters text, the system automatically detects the emotion (\textcircled{1}) and contextual keywords (\textcircled{2}) from the text, which are considered as key information used for psychological therapy assessment~\cite{ehrenreich2007role}. Specifically, the system recommends (i) one most likely representative emotion and (ii) up to three most salient contextual keywords. In this process, to ensure user agency, the user may also enter their own customized keywords.

For recommending the representative emotion, pre-trained model classifies texts in one among six emotions (\textit{anger}, \textit{sadness}, \textit{happiness}, \textit{surprise}, \textit{fear}, and \textit{distress}). Specifically, this taxonomy follows six basic emotions suggested by Ekman~\cite{ekman1999basic}, except for ‘disgust’ altered by ‘distress’ to cover as many topics in OMHCs as possible. Since we deployed model structure to be generalizable to other emotion sets as well, we believe our emotion categorization can easily be altered to other sets of emotions. Detailed training processes and the usage of models are described in Section~\ref{model}.

\subsubsection{Reacting to the post}

To assist support providers to react empathetically, we decided to use EPITOME~\cite{sharma2020computational}, a text-based empathic reaction framework. Devised through the collaboration with psychologists and synthesizing existing literature, EPITOME is conceptualized as follows: (i) emotional reaction (expressing emotional phrases of their own), (ii) interpretation (delivering understandings based on seekers' writing), and (iii) exploration (exploring information not stated in the post). In this process, by leveraging emotion detected and saved from the writing process, we decided to support (i) emotional reaction as a form of \textit{trigger}, with the rest of the processes guided by the \textit{prompts} (Figure \ref{fig:teaser}-(c)).

Specifically, we designed our system to recommend providers with three representative phrases to start their reaction with (trigger; \textcircled{3}). To collect the phrases suitable for users to start with, we first recruited a mental health counselor from the university, who has more than 2 years of counseling experience. Then, we asked them to recommend three short phrases of emotional reaction targeted to each emotion, along with two phrases that can be used for every negative emotion. After gathering all the responses, we designed the system to randomly choose and present two among three targeted phrases based on the detected emotion of the post, and one between two generalized emotions.

Once the user initiates writing a comment by selecting one of the recommended phrases (emotional reaction), they can start entering their own feedback regarding the post based on the prompts offered (\textcircled{4}). Here, the system offers prompts that guide users to (i) interpret the situation with their own words (interpretation) and then (ii) throw a question about the post (exploration), consecutively. Each parenthesis above indicates the element of EPITOME~\cite{sharma2020computational}, the empathic reaction framework we decided to utilize.

\subsubsection{Exploring posts}

In our interface, posts are sorted by time by default. Here, filtering tags located on top of the interface can be used to filter posts including specific emotions or contexts. Each of the emotion/context tag collections consists of tags collected from entire posts and once tags are selected, the system only shows posts containing more than one selected tag until entire tags are deselected.

\subsection{Model Selection} \label{model}


\subsubsection{Emotion detection}

For building an emotion classification model, we utilized ELECTRA~\cite{clark2020electra}, a state-of-the-art language training model, with fine-tuning targeted to the main language of our interface (Korean)~\cite{kim2020lmkor}. Then, we trained using Korean emotional dialogue corpus~\cite{aihub}, a dataset used for the training of the model that includes 270K transcribed sentences of emotional dialogues. Among various emotions available for the training, data that fall under six basic emotions that we previously defined were used to train the model. As such, we obtained 4.6K sentences and split them into approximately $8:1$ ratio for training and validation purposes. Then, we trained the model with a learning rate of 1E-5 for 3 epochs with the batch size of $N=16$, where we gained the validation set accuracy of 0.7083.

\subsubsection{Keyword extraction}

For the contextual keyword extraction, we used KRWordRank~\cite{kim2014kr}, a WordRank-based unsupervised Korean word extraction method. Once the keyword extraction begins, the system triggers the WordRank-based algorithm~\cite{mihalcea2004textrank} by default to extract the maximum of 3 keywords from the posting. In this process, if the algorithm fails to extract the keyword, the system performs lemmatization using Okt~\cite{okt}, a Korean text processor specialized in community postings. For both WordRank and lemmatization, we predefined unnecessary stopwords (e.g. today, me) and have those tasks drop stopwords in order to enhance the quality of recommendations.

Our system was developed as a mobile app targeted to iOS environment. The server for our detection models was deployed on AWS EC2, and user responses are programmed to be saved on Google Firebase with an anonymous~ID.
\section{Preliminary User Study}

The goal of this study is to identify the feasibility and possible enhancements of our proposed AI-assisted emotional support workflow. As such, we ran a preliminary user study with 10 participants using the system we designed.

\subsection{Methods}

\subsubsection{Participants}
From two major college online communities in South Korea, we recruited participants who self-reported that they often required emotional support and share their emotions in an online medium such as OMHCs or social media. Since we designed and deployed our system targeted to mobile settings, we screened users with the self-reported daily usage of smartphones of at least 4 hours on average for ensuring their familiarity with mobile settings. In addition, people who are currently in the process of mental therapy were excluded to avoid conflict with it. As such, 10 participants were recruited (\textit{M$_{\text{age}}$} = 22.6, \textit{SD$_{\text{age}}$} = 1.84; 6 female).

\subsubsection{Study procedure}
\begin{figure*}[h!]
    \centering
    \includegraphics[width=.9\linewidth]{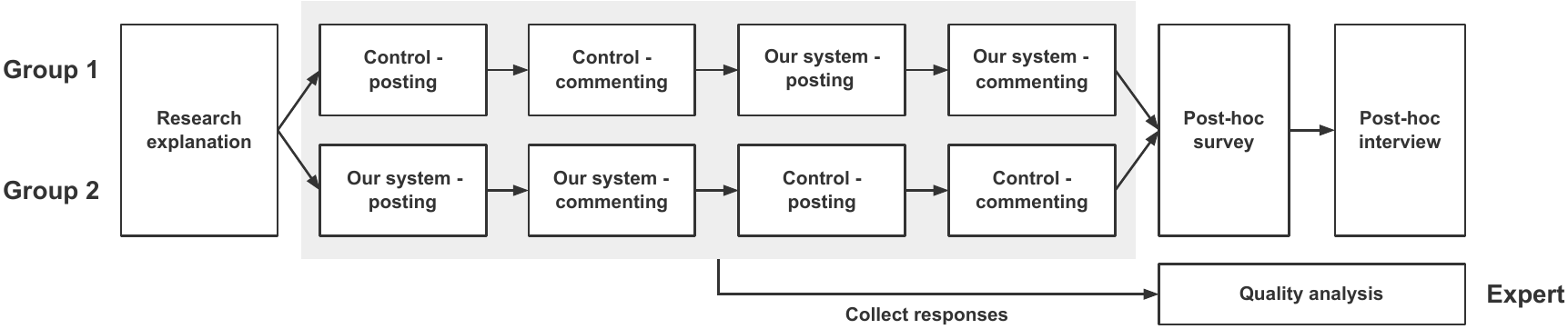}
    \caption{Research procedure of our study. 10 participants were randomly assigned to two conditions and used our/control interfaces, followed by the survey and interview sessions. Once the study for 10 participants are all complete, their responses are passed to mental health counselor to rate their quality}
    \label{fig:study-procedure}
\end{figure*}

In order to compare our system with the existing mental health community settings, we designed a \textit{control} interface, whose interface is identical except for the recommendation functionalities offered while users write, explore, and comment on the posts. Then, we ran a within-subject design study where 5 participants were assigned to use \textit{control interface} and \textit{our interface} consecutively, and reversed order for the 5 rest participants, to control the order effect. The overall procedure is illustrated in Figure~\ref{fig:study-procedure}.

First, we briefly introduced the research objectives and process, and the participants were given a link that guides them to install our prototype application. In each condition, the participants were requested to (i) write two new posts about anxious situations they often faced in daily life, and (ii) write reactions to two of others’ posts. For users to follow (ii), we showed participants with pre-defined posts by crawling existing posts from Korean OMHC after removing all the sensitive information. None of the participants’ data was reused for the purpose of another participant's study.

Each participant was then asked to complete the survey (5-point Likert scale; 1 = strongly disagree, 5 = strongly agree) regarding (i) the ease of using the interface for both our and control system and (ii) satisfaction with the assist of AI for each interaction. Then, we ran a semi-structured interview session to collect their overall experiences of using our system and its future enhancements. Each interview response was transcribed and later open-coded.

Once the study procedures for every participant were complete, each of the participants was compensated with 15,000 KRW (12 USD equivalent). Then, we passed their response data to university mental health counselor and had them evaluate posts and comments based on the following metric on a 5-point Likert scale, respectively: \textit{(i) how much does the post draw emotionally supportive reaction?} \textit{(ii) how much does the comment show empathy toward the post?} In this process, to mitigate bias stemming from the interface design, we had the counselor evaluate them only by viewing texts without showing any interface element.

\subsubsection{Privacy, ethics, and disclosure}

Our system asks users to take note of their emotional contexts, which are later passed to the mental health counselor for the purpose of evaluation. Thus, we considered it particularly important to strictly ensure their privacy during the study. As such, we formulated and followed the following precautionary steps: (i) we had our study approved by the university IRB and strictly followed the approved procedure, (ii) all the privacy-sensitive information (e.g., name, affiliation) in transcription and user-provided data was masked once collected.

\subsection{Results}


\begin{figure}[h!]
    \centering
    \includegraphics[width=\linewidth]{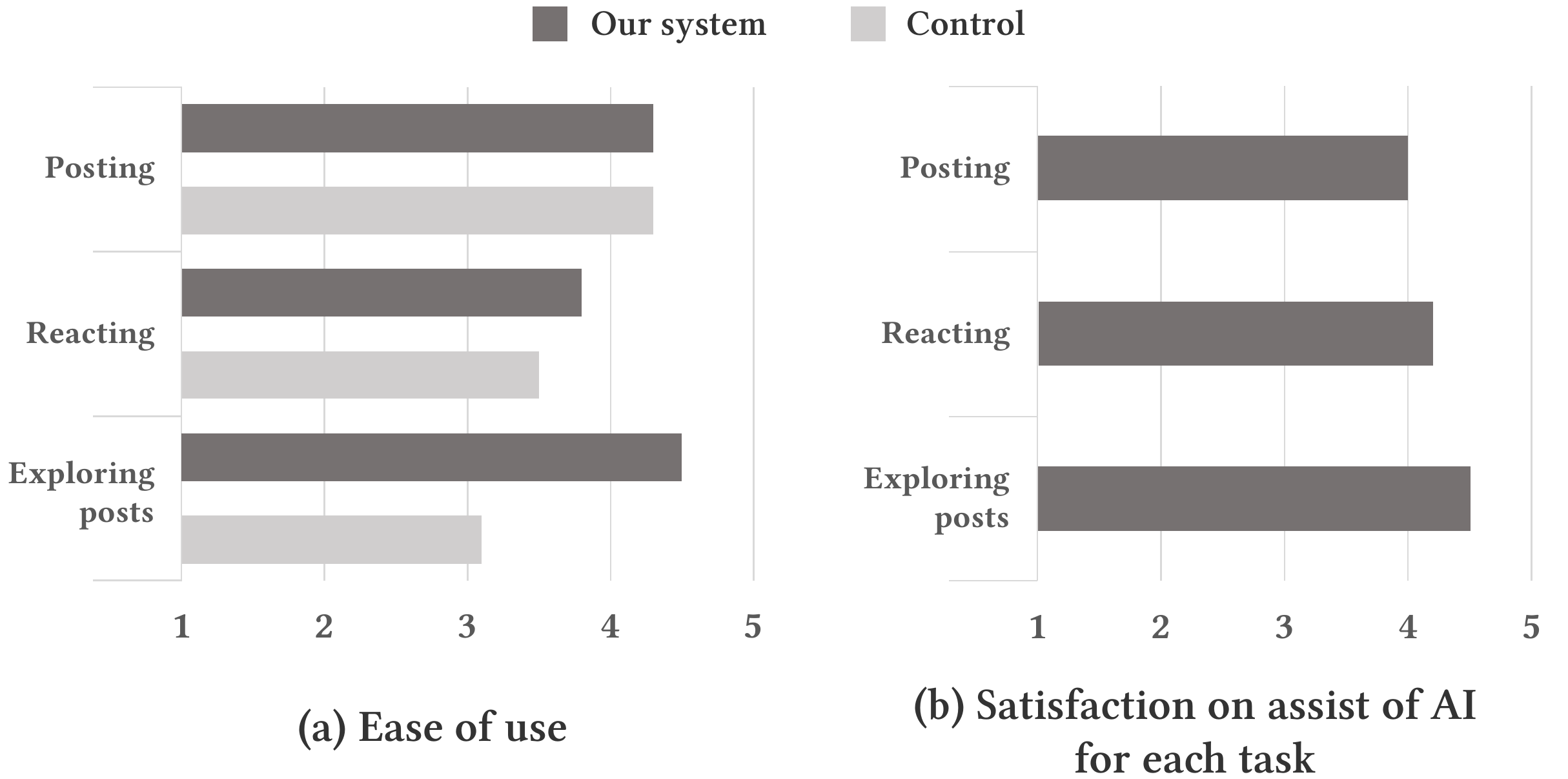}
    \caption{Survey of our study. (a) Users find it less burdensome to use our interface compared to the control interface, except for the posting. (b) AI-assisted posting (i.e., keyword recommendation), reacting (i.e., showing triggers/prompts to assist reaction), and exploring posts (filtering posts with keywords collected from posts) had the average satisfaction score of 4.0 or above}
    \label{fig:survey}
\end{figure}

\subsubsection{Writing a post}

Our system induces users to view the detected emotion and contextual keywords and lets them iterate writing based on them. Regarding such an approach, participants offered several positive feedback:

\begin{itemize}
    \item \textbf{Clarifying emotion and circumstance one faced}: First of all, participants responded that our system helped them become aware of the emotion and circumstances they face, before writing down the post: \textit{“When I was pretending as if I was emotionally stable, the system elicited my emotion through leading to iterate over, which led me to better figure out myself.”}~(P10)
    
    \item \textbf{Help express their thoughts as a form of text}: In addition to clarifying the emotion and circumstance they faced, such iterative processes of getting feedback from AI and re-writing texts were also reported to help them the process of converting such thoughts into writing. Since our system detects and presents the emotion/contextual keywords and keeps updating once the post is edited, they reported that they kept making their posts more machine-detectably concrete: \textit{“The system led me to keep in mind that I should show emotion and contexts in my post, leading to writing in a more readable and structured way.”}~(P7)
    
    However, analysis from the mental health counselor shows that such concreteness was not directed to the increase of the level of inducing empathetic reaction ($t=-0.1787$, $p=0.43$). Thus, further design consideration is required to lead the concreteness of the post to inducing an empathetic reaction.
    
    
    \item \textbf{Emotional reliance toward the system}: Interestingly, some participants reported that they could emotionally rely on the system. In other words, by viewing that our system actively suggests the emotion/contextual keywords as well as the way of reaction, they felt that the system listened to and would help them in terms of their emotional challenges. They anticipated that this would make them feel relieved in real-life emotional challenges: \textit{“Seeing the recommendations, I felt like the system would solve the emotional issue I faced.”}~(P4)
\end{itemize}

On top of the potential of our post-writing interface, we also identified several limitations of our system as follows:

\begin{itemize}
    \item \textbf{Increased awareness of others}: One participant showed the concern of being more aware of other people. Since our interaction unit is emotion and contextual keywords curated the seeker, the participant reported they became more aware of the others while writing the post: \textit{“I wrote carefully because I thought other people will look at my tags and consume them.”}~(P1)
    
    \item \textbf{Reluctance toward a feeling of being ‘diagnosed’ by AI}: In addition to the increased awareness toward others, one user pointed out the feeling of being ‘diagnosed’ by the emotion suggested by our system. Specifically, by having their emotion detected by the model, the system may require the user to become unexpectedly sober in terms of their emotion: \textit{“When writing a post, unintentional ‘diagnosis’ might be made through processes such as emotional judgment, so I thought the room for own deliberation had been decreased.”}~(P10)
\end{itemize}

\subsubsection{Reacting to the post}

From the analysis of mental health counselor, comments from our system ($M=4.55$, $SD=0.51$) presented significant improvement in terms of empathy compared to comments from control ($M=3.25$, $SD=0.92$; $t=-4.4316$, $p<0.0001$). In addition to possible attributes that might have led to such enhancement, we describe several benefits that our interface gave to the users:

\begin{itemize}
    \item \textbf{Opportunity to learn how to empathize with others}: Along with the comment trigger for emotional reaction, our system provides prompts that induce providers to include interpretation and exploration of the post, which results in the lowered burden of commenting. This acted as a ‘learning tool’ which helped people, who originally had difficulty, learn how to react empathetically: \textit{“I had been unfamiliar with how to react to such harsh feelings, but using this system, I could learn it through the trigger and guidance of the system.”} (P2)
    
    \item \textbf{Fostering a healthy community}: Considering the vulnerability of users in OHMCs, it is crucial to prevent harmful feedback or comments to keep OMHCs safe~\cite{chancellor2016post}. In this study, participants responded that the AI-assisted reaction process of our system has the potential of inducing a healthy feedback process and maintaining the community safer: \textit{“By viewing how the technology (triggers/prompts for reaction) offers candidates for neutralized expressions, I thought that people in (online mental health) communities may reduce their interpersonal conflicts.”}~(P5)
     
    \item \textbf{Improved understanding towards posts}: Participants reported that explicitly revealed emotional/contextual information enabled them to easily relate to a post and also helped deepen their understanding about the post: \textit{“Starting by understanding overall contexts of the post, I could better understand and empathize the post.”}~(P7)
\end{itemize}

At the same time, we could also gain several challenges and enhancements regarding the reaction process:

\begin{itemize}
    \item \textbf{Limiting the style of overall supports}: Although the commenting process of our system has elicited a positive range of responses for most of the participants, one participant (P9) expressed concern about the process of ‘regularizing’ reactions. They worried if our reaction-support method would make the overall comments in a too empathetic way, leading to reduce benefits from the reaction with opposite views or emotions (e.g., energetic): \textit{“When I have a hard time, I sometimes want to meet a lively person to get better (...) I was worried if most of the reactions (in our interface) would overflow only with the empathy toward sadness.”}~(P9)
    
    \item \textbf{Burden of comment-writing process}: The survey result for the AI-assisted commenting system was shown positive. Still, two of the participants (P7, P9) reported that the interactions patterns might be burdensome in terms of their long-term usage since it requires users to follow the prompts: \textit{“Overall load of clicking one of the triggers and following prompts was pretty high for me.”}~(P7)
\end{itemize}

\subsubsection{Exploring posts}

Lastly, participants showed unanimity on the efficiency of the filtering system in terms of exploring posts. Taking into account that the tag-based filtering system is commonly found in online communities, such positive feedback implies the feasibility of integrating our AI-driven keyword suggestions into the existing filtering system in OMHCs: \textit{“I usually look for articles that have similar situations/emotions to me (...) it was convenient curating the posts that have similar emotions like mine.”}~(P9)
\section{Discussion \& Limitation}

Throughout the study, we could identify the feasibility of AI-assisted writing processes in supporting interpersonal interactions in OMHCs, as well as its feasibility of forming a healthier community environment. Specifically, leveraging AI-driven emotion/contextual keyword elicitation was reported to induce seekers to clarify expression for AI to better understand, ultimately assisting the writing process to be more concrete. However, from the expert analysis, we realized that such concreteness did not necessarily lead to emotional support. Thus, further design iteration to connect from such concreteness to inducing empathetic reaction would be required.

Furthermore, our system was reported to assist the support providers by assisting them to learn how to react empathetically. Considering that such an empathetic reaction is a key to thriving OMHCs~\cite{kushner2020bursts}, we could see the feasibility of extending our work to long-term deployment settings. However, considering that some participants worried if our community setting might only be filled with monotonous, emotionally empathetic reactions, it is would also be beneficial to diversify the type of reactions (e.g., supporting energetic reactions) to enrich the community environment.

Still, our study presents several limitations. First, we conducted our study only with 10 participants. Thus, additional participants may be required in terms of generalizability. Second, this study was run in a lab-based setting, where participants were asked to follow designated actions. To collect a more lively experience of users in OMHCs aided by the AI-assisted writing process, a deployment study in the wild might be needed.
\section{Conclusion}

In this work, we presented an AI-infused writing workflow in OMHCs that penetrates from the seeker's posting to the provider's reaction. Leveraging AI models, our system recommends emotional/contextual information from postings and provides extra cues that help people to provide more empathetic responses based on their understandings toward seeker's personal experiences. Through the preliminary user study, we could identify the feasibility of AI-assisted emotional support processes in fostering healthy OMHCs, as well as their enhancements.

\begin{acks}
This work has been led by the first author in the midst of his graduate school application season, and it would have been impossible to finalize this work without the assistance from many people on his applications. Thus, we would like to genuinely thank Jaeyoon Song, Wesley Deng, Katelyn Morrison, Joanne Ma, Soomin Kim, and many others for their invaluable help on Donghoon's road to his graduate studies.

This work was supported by SNU Undergraduate Research Program through the Faculty of Liberal Education, Seoul National University (2021-05).
\end{acks}

\bibliographystyle{ACM-Reference-Format}
\bibliography{10-references}

\end{document}